\documentclass[twocolumn,prl,showpacs,floatfix]{revtex4}

\usepackage{graphicx}
\usepackage{dcolumn}
\usepackage{bm}
\usepackage{amsmath}

\begin{document}

\title{
Electron and boson clusters in confined geometries: symmetry breaking in
quantum dots and harmonic traps
}

\author{Constantine Yannouleas}
\author{Uzi Landman}

\affiliation{School of Physics, Georgia Institute of Technology,
             Atlanta, Georgia 30332-0430}

\date{20 October 2005}

\begin{abstract}
We discuss the formation of crystalline electron clusters in semiconductor 
quantum dots and of crystalline patterns of neutral bosons in harmonic traps. 
In a first example, we use calculations for two electrons in an elliptic quantum
dot to show that the electrons can localize and form a 
molecular dimer. The calculated singlet-triplet splitting ($J$) 
as a function of the magnetic field ($B$) agrees with 
cotunneling measurements with its behavior reflecting the effective dissociation 
of the dimer for large $B$. Knowledge of the dot shape and of $J(B)$ allows 
determination of the degree of entanglement. In a second example,
we study strongly repelling neutral bosons in two-dimensional harmonic traps.
Going beyond the Gross-Pitaevskii (GP) mean-field approximation, we show that 
bosons can localize and form polygonal-ring-like crystalline 
patterns. The total energy of the crystalline phase saturates in contrast to the
GP solution, and its spatial extent becomes smaller than that of the GP 
condensate.
\end{abstract}


\maketitle



Explorations of the size-dependent evolution of the properties of materials 
are at the frontier of modern condensed matter and materials research. 
Indeed, investigations of clusters containing a finite, well-defined, 
number of elementary building units (atoms, molecules, electrons or other 
elementary constituents), allow investigations of the transition from the 
atomic or molecular regime to the finite nano-aggregate domain, and 
ultimately of the convergence with increasing size to the condensed phase, 
extended system, category. Moreover, investigations of clusters provide 
opportunities for discovery of novel properties and phenomena that are 
intrinsic properties of finite systems, distinguishing them from bulk 
materials \cite{uzi}.

Commonly, studies of materials clusters 
involve atoms or molecules interacting through electrostatic or electromagnetic 
potentials, with the heavier nuclei being the ``structural skeleton'' 
and the much lighter electrons serving as the ``glue'' that binds the 
atoms together.
In this article we focus on novel, somewhat exotic, types of clusters. In 
particular, we discuss clusters of electrons in man-made (artificial) quantum 
dots (QDs) created through lithographic and gate-voltage techniques at 
semiconductor interfaces, and clusters of neutral atoms in traps under 
conditions that may relate to formation of Bose-Einstein condensates (BECs). We 
illustrate that these cluster systems reveal interesting emergent physical 
behavior arising from spontaneous breaking of spatial symmetries; symmetry
breaking (SB) is defined as a circumstance where a lower energy solution of the
Schr\"{o}dinger equation is found that is characterized by a lower
symmetry than that of the Hamiltonian of the system. Such SB is exhibited 
through the formation of clusters of {\it localized\/} electrons (often called 
Wigner molecules, WMs) in two-dimensional (2D) QDs  (see Fig.\ 1 and 
in particular our discussion of two-electron WMs below \cite{yl70}).
Symmetry breaking is also manifested in the transition \cite{roma}, induced by 
increasing the interatomic repulsive interaction strength, of the BEC state of 
neutral atoms confined by a parabolic 2D trap to a crystalline cluster state.\\
~~~~~~~\\

\begin{figure}[b]
\centering\includegraphics[width=6.5cm]{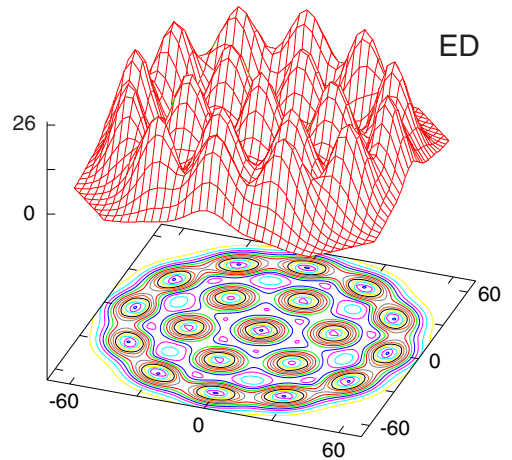}
\caption{
Unrestricted Hartree-Fock electron density in a parabolic QD for $N=19$ 
electrons and $S_z=19/2$, exhibiting
breaking of the circular symmetry at $R_W=5$ and zero magnetic field.
Remaining parameters are: parabolic confinement, $\hbar \omega_0=5$ meV;
effective mass $m^*=0.067 m_e$.
Distances are in nanometers and the electron density in $10^{-4}$ nm$^{-2}$.
}
\end{figure}

\noindent
{\bf Two-electron Wigner molecules.}
Electron localization leading to formation of molecular-like structures 
[the aforementioned Wigner molecules] within a {\it single circular\/} 
two-dimensional quantum dot at zero magnetic field ($B$) has been
theoretically predicted to occur \cite{yl11,grab,yl12,boni,mat,yl2p,yl2p2,yl13}, 
as the strength of the 
$e$-$e$ repulsive interaction relative to the zero-point energy increases, as 
expressed through an inceasing value of the Wigner parameter $R_W$, defined as 
$R_W=Z^2e^2/(\hbar \omega_0 l_0)$, with
$l_0=\sqrt{\hbar/(m\omega_0)}$ being the characteristic harmonic-oscillator
length of the confining potential. (The subscript $W$ in the case of a Coulomb 
force stands for ``Wigner'', since the confined clusters of localized electrons
may be viewed as finite-size precursors of the bulk Wigner crystal 
\cite{wig}.)
Formation of such ``molecular structures'' is a manifestation of
spontaneous symmetry breaking associated with a quantum phase transition, 
occuring at zero magnetic field for circular 2D QDs for $R_W \geq 1$
and involving the crossover from a liquid-like state to a crystalline one.
This crossover, described using symmetry breaking \cite{yl11}, was confirmed in 
other studies \cite{grab,boni,mat,yl2p2} using a variety of methods. 
The degree of electron localization, that underlies the appearance of
crystalline patterns, has been described \cite{yl11} as a progression from
``weak'' to ``strong'' Wigner molecules as a function of increasing $R_W$
(or equivalently decreasing density). In high magnetic 
fields, the rotating electron molecules exhibit magic angular momenta 
\cite{mak,y1,y2,yl4} corresponding to fractional quantum Hall effect (FQHE) 
fillings. This has led \cite{y1,y2,yl4} to the derivation of an analytic trial 
wave function that provides a better description of the finite-size analogs of 
the FQHE in comparison with the Laughlin \cite{laug} and composite-fermion 
\cite{jain} wave functions.

Because of the finite size, Wigner molecules are expected to show new behavior
that differs from the classical Wigner crystal familiar from Solid State
Physics. The limit of a classical Wigner crystal is expected to be reached for a
higher number of electrons $N$ and very large $R_W$ \cite{yl11}. 
In the following we use the term ``Wigner molecule'' even in the case of
only two localized electrons. For this case the WM exhibits close analogies to an
H$_2$ natural molecule, as described below.

Here we focus on a two-electron $(2e)$ WM, in light of the current
experimental effort \cite{taru,eng} aiming at implementation of a
spin-based \cite{burk} solid-state quantum logic gate that employs two coupled 
one-electron QDs (double dot). 
We present an exact diagonalization (EXD) and an approximate
(generalized Heitler-London, GHL) microscopic treatment for two electrons in a
{\it single\/} elliptic QD specified by the 
parameters of a recently investigated experimental device \cite{marc}. While 
formation of Wigner molecules in circular QDs requires weak confinement (that is
small $\omega_0$ in the expression for $R_W$ given above), 
and thus large dots of lower densities (so that the interelectron repulsion 
dominates), we show that formation of such WMs is markedly enhanced in highly 
deformed (e.g., elliptic) dots due to their lower symmetry.
The calculations provide a good description
of the measured $J(B)$ curve (the singlet-triplet splitting) when screening 
\cite{kou2,hall} due to the metal gates and leads is included (in addition to the
dielectric constant of the semiconductor, GaAs). In particular, our results
reproduce the salient experimental findings pertaining to the vanishing of 
$J(B)$ for a finite value of $B \sim 1.3 $ T [associated with a change in sign 
of $J(B)$ indicating a singlet-triplet (ST) transition], 
as well as the flattening of the $J(B)$ curve after the ST crossing.
These properties, and in particular the latter one, are related 
directly to the formation of an electron molecular dimer and its effective 
dissociation for large magnetic fields. 
The effective dissociation of the electron dimer is most naturally described 
through the GHL approximation, and it is fully supported by the more accurate,
but physically less transparent, EXD.

Of special interest for quantum computing \cite{burk} is the degree of 
entanglement exhibited by the two-electron molecule in its singlet state.
Entanglement is a purely quantum mechanical phenomenon in which the quantum 
state of two or more objects cannot be described independently of each other,
even when the individual objects are spatially separated. The highest degree
of entanglement occurs at full separation, as discussed in the celebrated 
EPR (Einstein, Podolsky, and Rosen) paper \cite{epr}. Electrons confined in a
quantum dot are not necessarily spatially separated from each other, and
consequently their degree of entanglement may be lower than the
maximal one, as shown by us below.
 
Here, in relation to the microscopic calculations, we investigate two different 
measures of entanglement. The first, known as the concurrence 
(${\cal C}$) for two {\it indistinguishable\/} fermions \cite{schl,loss}, 
has been used in the analysis of the experiment in Ref.\ \cite{marc}
(this measure is related to the operational cycle of a two-spin-qubit
quantum logic gate \cite{schl,loss}).
The second measure, referred to as the von Neumann entropy (${\cal S}$) for 
{\it indistinguishable\/} particles, has been developed in Ref.\ \cite{you}
and used in Ref.\ \cite{zung}.
We show that the present {\it wave-function-based\/} methods, in conjunction with
the knowledge of the dot shape and the $J(B)$ curve, enable theoretical 
determination of the degree of entanglement, in particular for the elliptic QD 
of Ref.\ \cite{marc}. The increase in the degree of entanglement (for both
measures) with stronger magnetic fields correlates with the dissociation
of the $2e$ molecule. This supports the experimental assertion \cite{marc} that 
cotunneling spectroscopy can probe properties of the electronic wave function of
the QD, and not merely its low-energy spectrum. Our methodology can be 
straightforwardly applied to other cases of strongly-interacting devices, e.g., 
double dots with strong interdot-tunneling.\\
~~~~~~~\\

\noindent
{\bf Clusters of neutral bosons in harmonic traps.}
Bose-Einstein condensates (BECs) in harmonic traps \cite{corn,corn2}
are normally associated with weakly interacting neutral atoms, and their physics
is described adequately by the Gross-Pitaevskii (GP) mean-field theory 
\cite{dal}. Lately, however, experimental advances in controlling the 
interaction strength \cite{cor,grei,par,wei} permit the production of novel 
bosonic states in the regime of strong interparticle repulsions. 
Theoretical efforts motivated by this capability include studies of the 
Bose-Hubbard model \cite{jak,mot}, and investigations about the 
``fermionization'' limit of an one-dimensional (1D) gas of trapped impenetrable
bosons \cite{gir2,dunj,ast}, often referred to as the Tonks-Girardeau 
(TG) regime \cite{gir2,gir}. Here we address the problem 
of strongly repelling (impenetrable) bosons in higher dimensions.
In particular, we discuss 2D interacting bosons in a {\it circular\/} harmonic 
trap, with the extension to 3D systems being straightforward. To this end,
we use computational methods that go beyond the GP method.

We explore the transition from a BEC (diffuse cloud) state to a
crystalline phase, in which the trapped localized bosons form 
crystalline patterns. At the mean-field level,
these crystallites are static and are portrayed directly in the
single-particle densities. After restoration of rotational symmetry, 
the single-particle densities are circularly symmetric, and thus the
crystalline symmetry becomes ``hidden''; however, it can be revealed in the
conditional probability distribution (CPD, anisotropic pair correlation), 
$P({\bf r},{\bf r}_0)$, which expresses the probability of finding a particle 
at ${\bf r}$ given that the ``observer'' (i.e., reference point) is riding on 
another particle at ${\bf r}_0$ \cite{yl2p,yl4}. 

\section{Methods}

\noindent
{\bf Two-electron quantum dot: Microscopic treatment.}
The Hamiltonian for two 2D interacting electrons is
\begin{equation}
{\cal H} = H({\bf r}_1)+H({\bf r}_2)+e^2/(\kappa r_{12}),
\label{ham}
\end{equation}
where the last term is the Coulomb repulsion, $\kappa$ is the
dielectric constant, and 
$r_{12} = |{\bf r}_1 - {\bf r}_2|$. $H({\bf r})$ is the
single-particle Hamiltonian for an electron in an external perpendicular
magnetic field ${\bf B}$ and an appropriate confinement potential.
When position-dependent screening is included, the last term in Eq.\
(\ref{ham}) is modified by a function of $r_{12}$ (see below). 
For an elliptic QD, the single-particle Hamiltonian is written as
\begin{equation}
H({\bf r}) = T + \frac{1}{2} m^* (\omega^2_{x} x^2 + \omega^2_{y} y^2)
    + \frac{g^* \mu_B}{\hbar} {\bf B \cdot s},
\label{hsp}
\end{equation}
where $T=({\bf p}-e{\bf A}/c)^2/2m^*$, with ${\bf A}=0.5(-By,Bx,0)$ being the
vector potential in the symmetric gauge. $m^*$ is the effective mass and
${\bf p}$ is the linear momentum of the electron. The second term is the
external confining potential; the last term is the Zeeman interaction with 
$g^*$ being the effective $g$ factor, $\mu_B$ the Bohr magneton, and ${\bf s}$ 
the spin of an individual electron. 

The GHL method for solving the Hamiltoninian (\ref{ham}) consists of two steps. 
In the first step, we solve selfconsistently the ensuing 
unrestricted Hartree-Fock (UHF) equations allowing for lifting of the 
double-occupancy requirement (imposing this requirement gives the 
{\it restricted\/} HF method, RHF).
For the $S_z=0$ solution, this step produces two single-electron 
orbitals $u_{L,R}({\bf r})$ that are localized left $(L)$ and right $(R)$ of the
center of the QD [unlike the RHF method that gives a single doubly-occupied 
elliptic (and symmetric about the origin) orbital]. 
At this step, the many-body wave function is a single Slater 
determinant $\Psi_{\text{UHF}} (1\uparrow,2\downarrow) \equiv 
| u_L(1\uparrow)u_R(2\downarrow) \rangle$ made out of the two occupied UHF 
spin-orbitals $u_L(1\uparrow) \equiv u_L({\bf r}_1)\alpha(1)$ and 
$u_R(2\downarrow) \equiv u_R({\bf r}_2) \beta(2)$, where 
$\alpha (\beta)$ denotes the up (down) [$\uparrow (\downarrow)$] spin. 
This UHF determinant is an eigenfunction of the projection $S_z$ of the total 
spin ${\bf S} = {\bf s}_1 + {\bf s}_2$, but not of ${\bf S}^2$ (or the parity
space-reflection operator). 

In the second step, we restore the broken parity and total-spin symmetries by 
applying to the UHF determinant the projection operator \cite{yl33,yl332} 
$P^{s,t}=1 \mp \varpi_{12}$, where the operator $\varpi_{12}$ interchanges the 
spins of the two electrons; the upper (minus) sign corresponds to the singlet. 
The final result is a generalized Heitler-London (GHL) two-electron wave function
$\Psi^{s,t}_{\text{GHL}} ({\bf r}_1, {\bf r}_2)$ for the ground-state singlet 
(index $s$) and first-excited triplet (index $t$), which uses
the UHF localized orbitals,
\begin{equation}
\Psi^{s,t}_{\text{GHL}} ({\bf r}_1, {\bf r}_2) \propto
\biglb( u_L({\bf r}_1) u_R({\bf r}_2) \pm u_L({\bf r}_2) u_R({\bf r}_1) \bigrb)
\chi^{s,t},
\label{wfghl}
\end{equation}
where $\chi^{s,t} = (\alpha(1) \beta(2) \mp \alpha(2) \beta(1))$ is the spin 
function for the 2$e$ singlet and triplet states.
The general formalism of the 2D UHF equations and of the subsequent restoration 
of broken spin symmetries can be found in Refs.\ \cite{yl13,yl33,yl332,yl44}.

The use of {\it optimized\/} UHF orbitals in the GHL is suitable for treating 
{\it single elongated\/} QDs. The GHL is equally applicable to double QDs with 
arbitrary interdot-tunneling coupling \cite{yl33,yl332}. In contrast,
the Heitler-London (HL) treatment \cite{hl} (known also as Valence bond), 
where non-optimized ``atomic'' orbitals of two isolated QDs are used, is 
appropriate only for the case of a double dot with small interdot-tunneling 
coupling \cite{burk}.

The orbitals $u_{L,R}({\bf r})$ are expanded in a real Cartesian 
harmonic-oscillator basis, i.e.,
\begin{equation}
u_{L,R}({\bf r}) = \sum_{j=1}^K C_j^{L,R} \varphi_j ({\bf r}),
\label{uexp}
\end{equation}
where the index $j \equiv (m,n)$ and $\varphi_j ({\bf r}) = X_m(x) Y_n(y)$,
with $X_m(Y_n)$ being the eigenfunctions of the one-dimensional oscillator in the
$x$($y$) direction with frequency $\omega_x$($\omega_y$). The parity operator
${\cal P}$ yields ${\cal P} X_m(x) = (-1)^m X_m(x)$, and similarly for $Y_n(y)$.
The expansion coefficients $C_j^{L,R}$ are real for $B=0$ and complex for finite
$B$. In the calculations we use $K=79$, yielding convergent results.

In the exact-diagonalization method, the many-body wave function is written as 
a linear superposition over the basis of non-interacting two-electron 
determinants, i.e.,
\begin{equation}
\Psi^{s,t}_{\text{EXD}} ({\bf r}_1, {\bf r}_2) =
\sum_{i < j}^{2K} \Omega_{ij}^{s,t} | \psi(1;i) \psi(2;j)\rangle,
\label{wfexd}
\end{equation}
where $\psi(1;i) = \varphi_i(1 \uparrow)$ if $1 \leq i \leq K$ and
$\psi(1;i) = \varphi_{i-K}(1 \downarrow)$ if $K+1 \leq i \leq 2K$ [and
similarly for $\psi(2,j)$].
The total energies $E^{s,t}_{\text{EXD}}$ and the coefficients
$\Omega_{ij}^{s,t}$ are obtained through a ``brute force'' diagonalization of
the matrix eigenvalue equation corresponding to the Hamiltonian in Eq.\ 
(\ref{ham}). The EXD wave function does not
immediately reveal any particular form, although, our calculations below
show that it can be approximated by a GHL wave function in the case of the
elliptic dot under consideration.\\
~~~~~~~\\

\noindent
{\bf Two-electron quantum dot: Measures of entanglement.}
To calculate the concurrence ${\cal C}$ \cite{schl,loss}, one needs a 
decomposition of the GHL wave function into a 
linear superposition of {\it orthogonal\/}
Slater determinants. Thus one needs to expand the {\it nonorthogonal\/} 
$u^{L,R}({\bf r})$ orbitals as a superposition of two other {\it orthogonal\/}
ones. To this effect, we write
$u^{L,R}({\bf r}) \propto \Phi^+({\bf r}) \pm  \xi \Phi^-({\bf r})$,
where $\Phi^+({\bf r})$ and $\Phi^-({\bf r})$ are the parity symmetric and 
antisymmetric (along the $x$-axis) components in their expansion given by 
Eq.\ (\ref{uexp}). 
Subsequently, with the use of Eq.\ (\ref{wfghl}), the GHL 
singlet can be rearranged as follows:
\begin{equation}
\Psi^{s}_{\text{GHL}} \propto
| \Phi^+(1\uparrow) \Phi^+(2\downarrow) \rangle - 
\eta |\Phi^-(1\uparrow)\Phi^-(2\downarrow) \rangle,
\label{rear}
\end{equation}
where the so-called interaction parameter \cite{loss}, $\eta=\xi^2$, is the 
coefficient in front of the second determinant.
Knowing $\eta$ allows a direct evaluation of the concurrence of the singlet
state, since ${\cal C}^s = 2\eta/(1+\eta^2)$ \cite{loss}. Note that 
$\Phi^+({\bf r})$ and $\Phi^-({\bf r})$ are properly normalized.
It is straightforward to show that $\eta=(1-|S_{LR}|)/(1+|S_{LR}|)$, where 
$S_{LR}$ (with $|S_{LR}| \leq 1$) is the overlap of the original 
$u^{L,R}({\bf r})$ orbitals.

For the GHL triplet, one obtains an expression independent of the 
interaction parameter $\eta$, i.e.,
\begin{equation}
\Psi^{t}_{\text{GHL}} \propto
| \Phi^+(1\uparrow) \Phi^-(2\downarrow) \rangle +
|\Phi^+(1\downarrow)\Phi^-(2\uparrow) \rangle,
\label{reart}
\end{equation}
which is a maximally (${\cal C}^t=1$) entangled state. Note that underlying the 
analysis of the experiments in Ref.\ \cite{marc} is a {\it conjecture\/} that 
wave functions of the form given in Eqs.\ (\ref{rear}) and (\ref{reart}) 
describe the two electrons in the elliptic QD.

To compute the von Neumann entropy, one needs to bring both 
the EXD and the GHL wave functions into a diagonal form (the socalled 
``canonical form'' \cite{schl2,you}), i.e.,
\begin{equation}
\Psi^{s,t}_{\text{EXD}} ({\bf r}_1, {\bf r}_2) =
\sum_{k=1}^M z^{s,t}_k | \Phi(1;2k-1) \Phi(2;2k) \rangle,
\label{cano}
\end{equation}
with the $\Phi(i)$'s being appropriate spin orbitals resulting from a unitary 
transformation of the basis spin orbitals $\psi(j)$'s [see Eq.\ (\ref{wfexd})]; 
only terms with $z_k \neq 0$ contribute. The upper bound $M$ can be 
smaller (but not larger) than $K$ (the dimension of the 
single-particle basis); $M$ is referred to as the Slater rank.
One obtains the coefficients of the canonical expansion from the fact that 
the $|z_k|^2$ are eigenvalues of the hermitian matrix $\Omega^\dagger \Omega$
[$\Omega$, see Eq.\ (\ref{wfexd}), is antisymmetric]. The von Neumann 
entropy is given by 
\begin{equation}
{\cal S} = -\sum_{k=1}^M |z_k|^2 \log_2(|z_k|^2)
\label{vonn}
\end{equation}
with the normalization $\sum_{k=1}^M |z_k|^2 =1$.
Note that the GHL wave functions in Eqs.\ (\ref{rear}) and (\ref{reart}) 
are already in canonical form, which shows that they always have a Slater rank 
of $M=2$. One finds ${\cal S}^s_{\text{GHL}} =
\log_2(1+\eta^2) - \eta^2 \log_2(\eta^2)/(1+\eta^2)$, and
${\cal S}^t_{\text{GHL}}=1$ for all $B$. For large $B$, the overlap 
between the two electrons of the dissociated dimer vanishes, and thus 
$\eta \rightarrow 1$ and ${\cal S}^s_{\text{GHL}} \rightarrow 1$.\\
~~~~~~~\\

\noindent
{\bf Neutral repelling bosons in harmonic traps: Gross-Pitaevskii equation.}
Mean-field symmetry breaking for bosonic systems has been discussed earlier in 
the context of two-component condensates, where each species is associated with 
a different space orbital \cite{esr1,esr2}. 
We consider here one species of bosons,
but allow each particle to occupy a different space orbital $\phi_i({\bf r}_i)$.
The permanent
$|\Phi_N \rangle = {\it Perm}[\phi_1({\bf r}_1), ..., \phi_N({\bf r}_N)]$
serves as the many-body wave function of the {\it unrestricted\/}
Bose-Hartree-Fock (UBHF) approximation. This wave function
reduces to the Gross-Pitaevskii form with the {\it restriction\/}
that all bosons occupy the same orbital $\phi_0({\bf r})$,
i.e., $|\Phi^{\text{GP}}_N \rangle =\prod_{i=1}^N \phi_0({\bf r}_i)$, and
$\phi_0({\bf r})$ is determined self-consistently at the restricted 
Bose-Hartree-Fock (RBHF) level via the equation \cite{esr3}
\begin{eqnarray}
 [ H_0({\bf r}_1) &+& (N-1) \int d{\bf r}_2 \phi^*_0({\bf r}_2)
V({\bf r}_1,{\bf r}_2) \phi_0({\bf r}_2)] 
\phi_0({\bf r}_1) \nonumber \\
&=& \varepsilon_0 \phi_0({\bf r}_1).
\label{gpe}
\end{eqnarray}
Here $V({\bf r}_1,{\bf r}_2)$ is the two-body repulsive interaction, 
which is taken to be a contact potential, 
$V_{\delta}= g\delta({\bf r}_1 -{\bf r}_2)$, for neutral bosons.
The single-particle hamiltonian is given by $H_0({\bf r}) = 
-\hbar^2 \nabla^2 /(2m) + m \omega_0^2 {\bf r}^2/2$, where $\omega_0$ 
characterizes the harmonic confinement.\\
~~~~~~~\\

\noindent
{\bf Neutral repelling bosons in harmonic traps: Symmetry breaking.}
We simplify the solution of the UBHF problem by considering explicit 
analytic expressions for the space
orbitals $\phi_i({\bf r}_i)$. In particular, since {\it the bosons must avoid 
occupying the same position in space in order to minimize their mutual
repulsion\/}, we take all the orbitals to be of the form of displaced 
Gaussians, namely, $\phi_i({\bf r}_i) = \pi^{-1/2} \sigma^{-1} 
\exp[-({\bf r}_i - {\bf a}_i)^2/(2 \sigma^2)]$. The positions ${\bf a}_i$  
describe the vertices of concentric regular polygons, with both the width
$\sigma$ and the radius $a=|{\bf a}_i|$ of the regular polygons  
determined variationally through minimization of the total energy
$E_{\text{UBHF}} = \langle \Phi_N | H | \Phi_N \rangle$
/$\langle \Phi_N | \Phi_N \rangle$, where 
$H = \sum_{i=1}^N H_0({\bf r}_i) + \sum_{i < j}^{N} 
V( {\bf r}_i,{\bf r}_j)$ is the many-body hamiltonian.
 
With the above choice of localized orbitals, the unrestricted permanent 
$|\Phi_N \rangle$ breaks the continuous rotational symmetry. However,
the resulting energy gain becomes substantial for stronger repulsion.
Controlling this energy gain (the strength of correlations) is the ratio 
$R_\delta$ between the strength of the 
repulsive potential and the zero-point kinetic energy. Specifically, for a 2D 
trap, one has $R_{\delta} = gm/(2\pi\hbar^2)$ for a contact potential.\\
~~~~~~~\\

\noindent
{\bf Neutral repelling bosons in harmonic traps: 
Restoration of broken symmetry.}
Although the optimized 
UBHF permanent $|\Phi_N \rangle$ performs exceptionally well regarding the 
total energies of the trapped bosons, in particular in comparison to the 
resctricted wave functions (e.g., the GP anzatz), it is still incomplete. 
Indeed, due to its localized orbitals, $|\Phi_N \rangle$ 
does not preserve the circular 
(rotational) symmetry of the 2D many-body hamiltonian $H$. Instead, it 
exhibits a lower point-group symmetry, i.e., a $C_2$ symmetry for $N=2$ and a 
$C_5$ one for the $(1,5)$ structure of $N=6$ (see below). As a result, 
$|\Phi_N \rangle$ does not have a good
total angular momentum. This paradox is resolved through a
post-Hartree-Fock step of {\it restoration\/} of broken symmetries 
via projection techniques \cite{yl44,yl3}, yielding a new wave 
function $|\Psi_{N,L}^{\text{PRJ}} \rangle$ with a definite angular 
momentum $L$, that is 
\begin{equation}
2 \pi |\Psi_{N,L}^{\text{PRJ}} \rangle = \int^{2\pi}_0 d\gamma
|\Phi_{N}(\gamma) \rangle e^{i\gamma L}, 
\label{wfprj}
\end{equation}
where $|\Phi_{N}(\gamma) \rangle$ is the original UBHF permanent having each 
localized orbital rotated by an azimuthal angle $\gamma$, with $L$ being  
the total angular momentum. The projection yields wave functions for a whole 
rotational band. Note that the projected wave function 
$|\Psi_{N,L}^{\text{PRJ}} \rangle$ in Eq.\ (\ref{wfprj}) may be regarded
as a superposition of the rotated permanents $|\Phi_{N}(\gamma) \rangle$,
thus corresponding to a ``continuous-configuration-interaction'' solution.

Here, we are interested in the projected ground-state ($L=0$) 
energy, which is given by
\begin{equation} 
E_0^{\text{PRJ}} = \langle \Psi_{N,0}^{\text{PRJ}}  | H | 
\Psi_{N,0}^{\text{PRJ}}  \rangle / 
\langle \Psi_{N,0}^{\text{PRJ}}  | \Psi_{N,0}^{\text{PRJ}}
\rangle.
\label{eprj0}
\end{equation}

\section{Results and discussion}

\begin{figure}[t]
\centering\includegraphics[width=6.5cm]{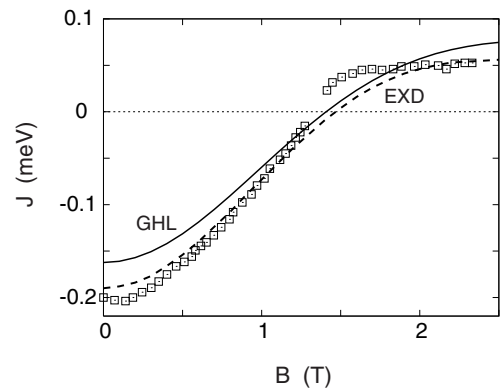}
\caption{
The singlet-triplet splitting $J=E^s-E^t$ as a function of the magnetic field
$B$ for an elliptic QD with $\hbar \omega_x=1.2$ meV and $\hbar \omega_y=3.3$ meV
(these values correspond to the device of Ref.\ \cite{marc}).
Solid line: GHL (broken-symmetry UHF + restoration of symmetries) results
with a coordinate-independent screening ($\kappa=22$).
Dashed line: EXD results with $\kappa=12.9$ (GaAs), but including screening with
a coordinate dependence according to Ref.\ \cite{hall} and $d=18.0$ nm 
(see text). The rest of the material parameters used are: 
$m^*$(GaAs)$=0.067 m_e$, and $g^*=0$ (see text).
The experimental measurements \cite{marc} are denoted by open squares.
Our sign convention for $J$ is opposite to that in Ref.\ \cite{marc}.
}
\end{figure}

\noindent
{\bf Two-electron quantum dot.}
To model the experimental elliptic QD device, we take, following Ref.\ 
\cite{marc}, $\hbar \omega_x=1.2$ meV and $\hbar \omega_y=3.3$ meV. 
The effective mass of the electron is taken as $m^*=0.067 m_e$ (GaAs). Since 
the experiment did not resolve the lifting of the triplet degeneracy caused by 
the Zeeman term, we take $g^*=0$. Using the two step method, 
we calculate the GHL singlet-triplet splitting 
$J_{\text{GHL}}(B)=E^s_{\text{GHL}}(B)-E^t_{\text{GHL}}(B)$ 
as a function of the magnetic field in the range $0 \leq B \leq 2.5$ T.
Screening of the $e$-$e$ interaction due to the metal gates and leads 
must be considered in order to reproduce the experimental $J(B)$ 
curve. This screening can be modeled, to first approximation, by a 
position-independent adjustment of the dielectric constant 
$\kappa$ \cite{kyri}. Indeed, with $\kappa=22.0$ (instead of the GaAs 
dielectric constant, i.e., $\kappa = 12.9$), good agreement with
the experimental data is obtained [see Fig.\ 2]. In particular, we note the 
singlet-triplet crossing for $B \approx 1.3$ T, and the flattening of the 
$J(B)$ curve beyond this crossing.

We have also explored, particularly in the context of the EXD treatment, 
a position-dependent screening using the functional form,
$(e^2/\kappa r_{12}) [1-(1+4d^2/r_{12}^2)^{-1/2}]$, 
proposed in Ref.\ \cite{hall}, with $d$ as a fitting parameter. 
The $J_{\text{EXD}}(B)$ result for $d=18.0$ nm is depicted in Fig.\ 2 
(dotted line), and it is in very good agreement with the experimental 
measurement.

\begin{figure}[t]
\centering\includegraphics[width=6.5cm]{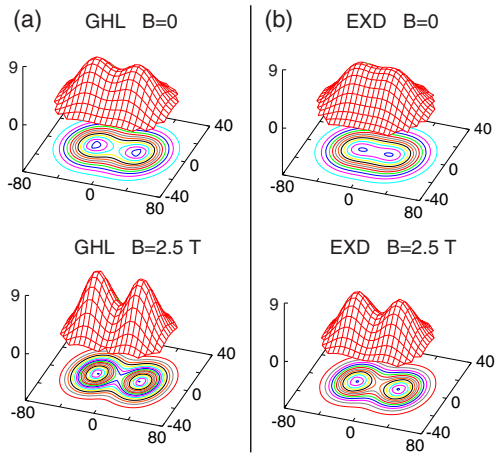}
\caption{
Total electron densities (EDs) associated with the singlet state of the
elliptic dot at $B=0$ and $B=2.5$ T.
(a) The GHL densities. (b) The EXD densities.
The rest of the parameters and the screening of the Coulomb interaction 
are as in Fig.\ 2.
Lengths in nm and densities in $10^{-4}$ nm$^{-2}$.
}
\end{figure}

The singlet state electron densities from the GHL and the EXD treatments
at $B=0$ and $B=2.5$ T are displayed in Fig.\ 3. These densities illustrate the 
dissociation of the electron dimer with increasing magnetic field. 
The asymptotic convergence (beyond the ST point) of the energies of the singlet 
and triplet states, i.e., [$J(B) \rightarrow 0$ as $B \rightarrow \infty$] is a 
reflection of the dissociation of the 2$e$ molecule, since the ground-state 
energy of two fully spatially separated electrons (zero overlap) does not 
depend on the total spin. 

In contrast, the singlet-state RHF electron densities fail to exhibit 
formation of an electron dimer for all values of $B$. This underlies the
failure of the RHF method to describe the behavior of the 
experimental $J(B)$ curve. In particular, $J_{\text{RHF}}(B=0)$ has the
wrong sign, while $J_{\text{RHF}}(B)$ diverges for high $B$ as is the case
for the RHF treatment of double dots (see Ref.\ \cite{yl33}).

For the GHL singlet, using the overlaps of the left and right orbitals, we find
that starting with $\eta=0.46$ $({\cal C}^s=0.76)$ at $B=0$, the interaction
parameter (singlet-state concurrence) increases monotonically to $\eta=0.65$ 
$({\cal C}^s=0.92)$ at $B=2.5$ T. At the intermediate value corresponding to the
ST transition ($B=1.3$ T), we find $\eta=0.54$ $({\cal C}^s=0.83)$.

Our $B=0$ theoretical results for 
$\eta$ and ${\cal C}^s$ are in remarkable agreement with 
the experimental estimates \cite{marc} of $\eta=0.5 \pm 0.1$ and 
${\cal C}^s=0.8$, which were based solely on conductance measurements below the 
ST transition (i.e., near $B=0$). 
We note that, for the RHF, ${\cal C}^s_{\text{RHF}}=0$, since a single 
determinant is unentangled for both the two measures considered here.

\begin{figure}[t]
\centering\includegraphics[width=6.5cm]{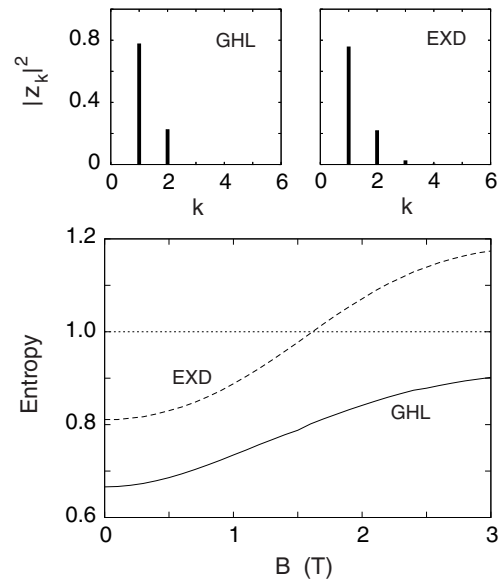}
\caption{
Von Neumann entropy for the singlet state of the elliptic dot as a function of 
the magnetic field $B$. Solid line: GHL. Dashed line: EXD. 
The rest of the parameters and the screening of the Coulomb interaction 
are as in Fig.\ 2. At the top, we show histograms for the $|z_k|^2$
coefficients [see Eq.\ (\ref{cano})] of the singlet state at $B=1.3$ T, 
illustrating the dominance of two configurations. Note the small third
coefficient $|z_3|^2=0.023$ in the EXD case.
}
\end{figure}

Since the EXD singlet has obviously a Slater rank $M > 2$, the definition
of concurrence is not applicable to it.  
The von Neumann entropy for the EXD singlet (${\cal S}^s_{\text{EXD}}$) is 
displayed in Fig.\ 4, along with that (${\cal S}^s_{\text{GHL}}$) of the GHL 
singlet. ${\cal S}^s_{\text{EXD}}$ and 
${\cal S}^s_{\text{GHL}}$ are rather close to each other for the entire
$B$ range, and it is remarkable that both 
remain close to unity for large $B$, although the maximum allowed mathematical 
value is $\log_2(K)$ [as aforementioned we use $K=79$, i.e., $\log_2(79)=6.3$];
this maximal value applies for both the EXD and GHL approaches. The saturation 
of the entropy for large $B$ to a value close to unity reflects the dominant 
(and roughly equal at large $B$) weight of two configurations in the 
canonical expansion [see Eq.\ (\ref{cano})] of the EXD wave function, which
are related to the two terms ($M=2$) in the canonical expansion of the GHL
singlet [Eq.\ (\ref{rear})]. This is illustrated by the histograms of the
$|z_k^s|^2$ coefficients for $B=1.3$ T at the top of Fig.\ 4.
These observations support the GHL approximation, which is
computationally less demanding than the exact diagonalization, and can be used
easily for larger $N$.\\
~~~~~~~\\

\begin{figure}[t]
\centering\includegraphics[width=6.5cm]{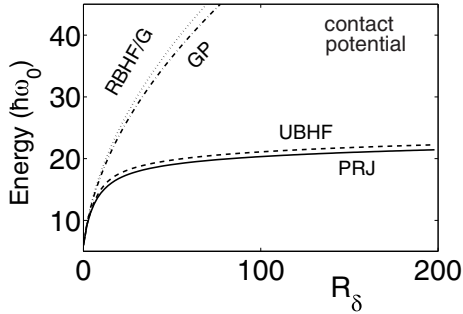}
\caption{
Total energies as a function of $R_{\delta}$ 
for various approximation levels, calculated for $N=6$ harmonically confined 
2D bosons in the (1,5) lowest-energy configuration.
Notation: RBHF/G - Restricted Bose-Hartree-Fock (RBHF) energy,
$E^{\text{Gauss}}_{\text{RBHF}}$ , with the common 
orbital $\phi_0({\bf r})$ approximated by a Gaussian centered at the 
trap origin; GP - the Gross-Pitaevskii energy; PRJ - the energy of the
symmetry-restored state obtained via projection of the (unrestricted)
UBHF state. Energies in units of $\hbar \omega_0$.}
\end{figure}

\noindent
{\bf Neutral repelling bosons in harmonic traps.}
In Fig.\ 5, we display  as a function of the
parameters $R_{\delta}$ the total energies for $N=6$ bosons
calculated at several levels of approximation. 
In both cases the lowest UBHF energies 
correspond to a (1,5) crystalline configuration, namely one boson is at 
the center and the rest form a regular pentagon of radius $a$. Observe
that the GP total energies are slightly lower than the 
$E_{\text{RBHF}}^{\text{Gauss}}$ ones; however, 
both exhibit
an unphysical behavior since they diverge as $R_{\delta} \rightarrow \infty$.
This behavior contrasts sharply with that of the unrestricted Hartree-Fock
energies, $E_{\text{UBHF}}$ and PRJ (see below), which saturate as 
$R_{\delta} \rightarrow \infty$; in fact, a value close to saturation is
achieved already for $R_{\delta}$ $\sim$ 10. 
We have checked that for all cases with 
$N=2 - 7$, the total energies exhibit a similar behavior. For a repulsive 
contact potential, the saturation of the UBHF energies is associated with the 
ability of the trapped bosons (independent of 
$N$) to minimize their mutual repulsion by occupying different positions in 
space, and this is one of our central results. 
For $N=2$, the two bosons localize at a distance $2a$ apart to form an 
antipodal dimer. For $N \leq 5$ the preferred UBHF crystalline arrangement is
a single ring with no boson at the center [usually denoted as $(0,N)$].
$N=6$ is the first case having one boson at the center [designated as 
$(1,N-1)$], and the (0,6) arrangement is a higher energy isomer.
The structural parameters (e.g., the width of the Gaussian orbitals and
the radii of the polygonal ring, calculated via the UBHF method, show a
saturation behavior similar to that illustrated above for the energy of the
system \cite{roma}. In contrast, the width of the condensate cloud (i.e., the 
GP solution) diverges with increasing repulsion strength ($R_\delta$).

The saturation found here for 2D trapped bosons interacting through 
strong repelling contact potentials is an illustration of the
``fermionization'' analogies that appear in strongly correlated systems
in all three dimensionalities. Indeed such energy saturation has been
shown for the TG 1D gas \cite{gir,gir2}, and has also been discussed for
certain 3D systems (i.e., three trapped bosons \cite{blum} and an infinite
boson gas \cite{heis}). Saturation of the energy
and the length of the trapped atom cloud (and thus of the interparticle
distance) has been measured recently for the 1D TG gas
(see in particular Fig.\ 3 and Fig.\ 4 in Ref.\ \cite{wei} and compare
to the similar trends predicted here for the 2D case in Fig.\ 5).

For $N=6$ 2D bosons, Fig.\ 5 shows that the $E_0^{\text{PRJ}}$ energies
share with the UBHF ones the saturation property for the case of a 
contact-potential repulsion. However, the projection brings further lowering 
of the total energies compared to the UBHF ones. 
(The projected ground state is always lower in energy than the original 
broken-symmetry one \cite{note43}.) 
Thus, for strong interactions 
(large values of $R_\delta$) the restoration-of-broken-symmetry step 
yields an excellent approximation of both the exact many-body wave function and
the exact total energy.

\begin{figure}[t]
\centering\includegraphics[width=6.5cm]{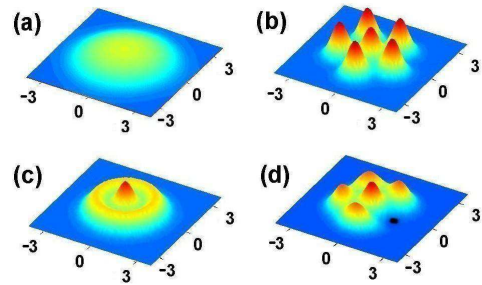}
\caption{
(a-c): Single-particle densities for $N=6$ 2D harmonically trapped neutral 
bosons with a contact interaction and $R_\delta=25$.
(a) The single-orbital self-consistent GP case. 
(b) The symmetry broken UBHF case (static crystallite).
(c) The projected case [symmetry-restored wave function, see Eq.\ (\ref{wfprj})].
The crystalline structure of the outer ring in this last case is ``hidden'',
but it is revealed in the conditional probability distribution \cite{yl2p,yl4} 
displayed in (d), where the observation point is denoted by a black dot (on the right). 
Lengths in units of $l_0$.
}
\end{figure}

The transformations of the single-particle densities (displayed in Fig.\ 6
for $N=6$ neutral bosons interacting via a contact potential 
and $R_\delta=25$) 
obtained from application of the successive approximations provide an 
illustration of the two-step method of symmetry breaking with subsequent 
symmetry restoration. Indeed, the GP 
single-particle density [Fig.\ 6(a)] is circularly symmetric, but the UBHF one 
[Fig.\ 6(b)] explicitly exhibits a (1,5) crystalline configuration. After 
symmetry restoration [Fig.\ 6(c)], the circular symmetry is
re-established, but the single-particle density is radially modulated unlike 
the GP density. In addition, the crystalline structure in the projected wave 
function is now hidden; however, it can be revealed through the use of the
CPD \cite{yl2p,yl4} [see Fig.\ 6(d)], which resembles the 
(crystalline) UBHF single-particle density, but with one of the humps 
on the outer ring missing (where the observer is located). In particular, 
$P({\bf r}_0, {\bf r}_0) \approx 0$ and the boson associated with the observer 
is surrounded by a ``hole'' similar to the exchange-correlation hole in 
electronic systems. This is another manifestation of the ``fermionization'' of
the strongly repelling 2D bosons. However, here as in the 1D TG case 
\cite{gir,gir2}, the vanishing of $P({\bf r}_0,{\bf r}_0)$ results from the
impenetrability of the bosons. For the GP condensate, the CPD 
is independent of ${\bf r}_0$, i.e., 
$P_{\text{GP}}({\bf r}, {\bf r}_0) \propto |\phi_0({\bf r})|^2$, reflecting
the absence of any space correlations.

It is of importance to observe that the radius of the BEC [GP case, Fig.\ 6(a)] 
is significantly larger than the actual radius of the 
strongly-interacting crystalline phase [projected wave function, Fig.\ 6(c)]. 
This is because the extent of the crystalline phase 
saturates, while that of the GP condensate grows with no bounds as $R_\delta
\rightarrow \infty$. Such dissimilarity in size (between the condensate and 
the strongly-interacting phase) has been also predicted \cite{dunj} for the 
trapped 1D Tonks-Girardeau gas and indeed observed experimentally \cite{wei}. 
In addition, the 2D single-particle momentum distributions for neutral bosons 
have a one-hump shape with a maximum at the origin (a behavior exhibited also 
by the trapped 1D TG gas). The width of these momentum distributions versus 
$R_\delta$ increases and saturates to a finite value, while that
of the GP solution vanishes as $R_\delta \rightarrow \infty$.

\section{Summary}

In this paper, we explored symmetry-breaking transitions predicted to occur
in confined fermionic and bosonic systems when the strength of the
interparticle repulsive interactions exceeds an energy scale that characterizes
the degree of confinement. For two electrons in an elliptic QD, we
predicted formation and effective dissociation (with increasing magnetic field)
of an electron dimer, which is reflected in the behavior of the computed
singlet-triplet splitting, $J(B)$, that agrees well (Fig.\ 2) with
measurements \cite{marc}.

Furthermore, we showed that, from a knowledge of the dot shape and 
of $J(B)$, theoretical analysis along the lines introduced here allows probing 
of the correlated ground-state wave function and determination of its degree of 
entanglement. This presents an alternative to the experimental study where
determination of the concurrence utilized conductance data \cite{marc}.
Such information is of interest to the implementation of spin-based 
solid-state quantum logic gates.

For the case of 2D trapped bosonic clusters, we found with increasing repulsive 
two-body interaction localization of the bosons in the trap, resulting in
formation of crystalline patterns made of polygonal rings; while we have focused 
here on repulsive contact interactions, similar results were obtained also for 
a Coulomb repulsion \cite{roma}.

These results provide the impetus for experimental efforts to access the
regime of strongly repelling neutral bosons in two dimensions. To this end we 
anticipate that extensions of methodologies developed for the recent realization
of the Tonks-Girardeau regime in 1D 
(using a finite small number of trapped $^{87}$Rb 
and optical lattices, with a demonstrated wide variation of $R_{\delta}$ 
from 5 to 200 \cite{par} and from 1 to 5 \cite{wei}) will prove most promising.
Control of the interaction strength via the use of the Feshbach resonance
may also be considered \cite{cor}.

This research is supported by the US D.O.E. (Grant No. FG05-86ER45234), and
NSF (Grant No. DMR-0205328).

\end{document}